\documentclass{desyproc}

\newcommand{\dpe}{D$\, \mathrm{I\!\!\!P}$E }
\usepackage{subfigure}
\begin{document}
%------------------------------------
\title{Exclusive central $\pi^+\pi^-$ production in CDF}

%for single authors the superscripts are optional
\author{{\slshape Michael Albrow$^1$, Artur Swiech$^2$, Maria Zurek$^2$}\\[1ex]
$^1$Fermilab. Wilson Road, Batavia, IL 60510, USA\\
$^2$Jagellonian University, Cracow, Poland\\
Presented at EDS Blois 2013: On behalf of the CDF Collaboration\\}

% if the proceedings are available online (e.g. at Indico)
% please enter the contribution ID or file_name below for the DOI
%\contribID{32}
%\contribID{smith\_joe}

% TO THE CONFERENCE EDITORS: 
% please update the following information      
% before sending the template to the authors
% \confID{800}  % if the conference is on Indico uncomment this line

\acronym{EDS'09} % if you want the Acronym in the page footer uncomment this line

\maketitle

\begin{abstract}
Using the Collider Detector at Fermilab, CDF, we have measured exclusive $\pi^+\pi^-$ production at $\sqrt{s}$ = 900 GeV and 1960 GeV.
The $\pi^+\pi^-$-pair is central, $|y| <$ 1.0, and there are no other particles detected in $|\eta| < 5.9$. We discuss the mass
spectrum, showing $f_0(980)$ and $f_2(1270)$ resonances,
$s$-dependence, $p_T$-dependence, and angular distributions.
\end{abstract}

\section{Introduction, CDF detector and data sets}
In Regge phenomenology, high mass single diffraction implies a non-zero triple-pomeron coupling , which in turn implies,
through the optical theorem, double pomeron exchange, \dpe,: $ p + p \rightarrow p(*) \oplus X \oplus p(*)$. Here $p$
means a proton or antiproton, the final state protons may be quasi-elastic or they may dissociate ($p(*)$), and $\oplus$ represents a
large rapidity gap $\Delta y \gtrsim 3$ with no hadrons. See Ref.\cite{acf} for a review. By ``exclusive" we mean that the central state
$X$ is simple and fully measured. At low masses, in the resonance region $M(X) \lesssim$ 3 GeV, \dpe is non-perturbative and QCD (or
QCD-inspired) calculations are challenging; there are new efforts by the Durham~\cite{durham} and Cracow~\cite{cracow} groups. The
quantum numbers of $X$ are restricted to be mostly $I^G J^{PC} = 0^+ \mathrm{even}^{++}$, so $s$-channel resonances $f_0(600), f_0(980),
f_2(1270), \chi_{c0}(3415)$ and $\chi_{c2}(3556)$ are allowed. Resonances with a high gluon content will be favored, especially
in comparison with $\gamma\gamma \rightarrow X$. For the $\chi_c$ and $\chi_b$ states perturbative calculations of $g+g
\rightarrow \chi_{c,b}$ are applicable, related to the very interesting channels $X = \gamma\gamma$~\cite{cdfgg} and $X =$ Higgs. So we have several
motivations: improving our understanding of the pomeron, meson (especially glueball) spectroscopy, and testing the QCD physics of
exclusive production (especially $\gamma\gamma$ and Higgs). 

The CDF detector at the Fermilab Tevatron is well-known. For this study we used data not only at the usual $\sqrt{s}$ = 1960 GeV, but
also at 900 GeV in a special run. We only used bunch crossings with a single interaction, i.e. no pile-up, and we required all the CDF
detectors, covering $-5.9 < \eta < +5.9$ to be empty, except for two oppositely-charged tracks and their corresponding calorimeter hits.
The trigger for these events was $\geq 2$ calorimeter showers with $E_T \gtrsim$ 0.5 GeV, with a veto on beam shower counter hits ($|\eta|
= 5.4 - 5.9$), Cherenkov luminosity counters ($|\eta|$ = 3.7 - 4.7) and forward calorimeters ($|\eta|$ = 1.32 - 3.64). We had 22M
(90M) triggers at $\sqrt{s}$ = 900 (1960) GeV. Off-line we required the central calorimeters ($|\eta|<$ 1.3) to be also
empty, apart from the trigger clusters.

\section{Exclusivity cuts, luminosity normalization, and event selection}

Importantly, we simultaneously recorded a large sample of 0-bias (bunch crossing) triggers. Dividing these into ``interaction" and
``non-interaction" samples, as in Ref.~\cite{cdfgg}, allowed us to determine the noise levels in all the detectors, and to measure the total visible
cross section $\sigma(vis)$, which is the inelastic cross section $\sigma(inel)$ times the fraction $f_{vis}$ of inelastic events with particles
in $|\eta| < $ 5.9, estimated to be 0.90$\pm$0.05 (0.85$\pm$0.05). At 1960 GeV $\sigma(vis)$ agreed with global fits; at 900 GeV we used
$\sigma(vis)$ to normalize our cross sections, as the luminosity counters were not calibrated. The total delivered luminosity at the two
energies was 0.056 (7.12) pb$^{-1}$. The effective ``no-pile-up" luminosity was 0.0435 (1.18) pb$^{-1}$, determined by counting empty
0-bias events as a function of the bunch luminosity.
Off-line we required exactly two well-measured opposite-charge tracks with $|\eta| < 1.3$ and $p_T >$ 0.4 GeV/c. The pair $X =
\pi^+\pi^-$ ($\pi$-masses assumed) was required to have $|y(\pi\pi)| <$ 1.0, and $M(\pi\pi) >$ 0.8 GeV to have acceptance down to $p_T = 0$.
We calculated the acceptance and efficiencies for the above fiducial region, and with the effective luminosity calculated the differential cross section
$d\sigma/dM(\pi\pi).dp_T(\pi\pi)$, assuming an isotropic (S-wave) $X \rightarrow \pi^+\pi^-$ distribution. 

\section{Results}

\begin{figure}
\centering
\includegraphics[width=125mm]{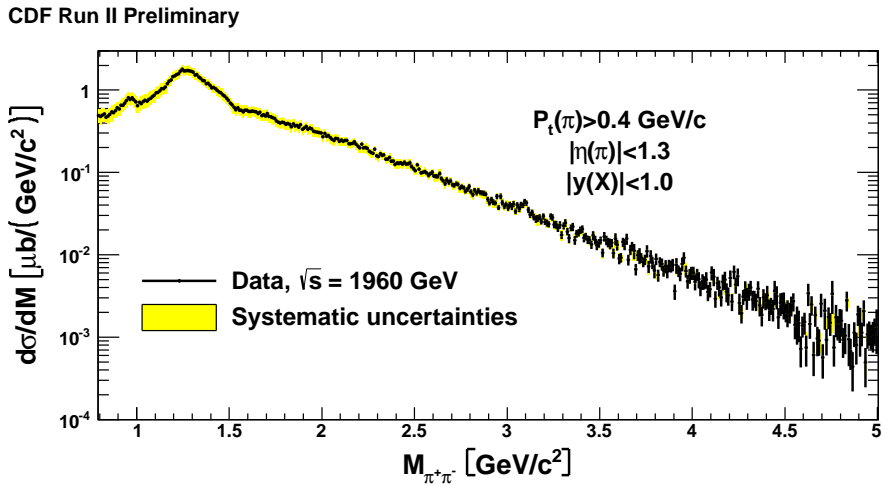}
\caption{Differential cross section $d\sigma/dM$ for two particles, assumed to be $\pi^+\pi^-$, in the stated kinematic region,
between two rapidity gaps $\Delta y > $4.6,  at $\sqrt{s}$ = 1960 GeV.}
\label{fig:Mass_Corr_1960}
\end{figure}

Fig.~\ref{fig:Mass_Corr_1960} shows the differential cross section integrated over $p_T(\pi\pi)$ as a function of $M(\pi\pi)$, 
and Fig.~2a shows the low mass region on a linear scale, and at both energies. A small $f_0(980)$ signal
is seen, and a dominant $f_2(1270)$ (also dominant in $\gamma\gamma \rightarrow \pi^+\pi^-$). A possible shoulder on the high
mass side ($f_0(1370)$?) is followed by a distinct change of slope at 1500 MeV, which was also seen at lower
energies~\cite{afs}. While the cross section shapes are similar at the two energies, they differ in detail as seen in the
ratio plot Fig.~2b. In addition to any $s$-dependence of the $p \oplus \pi^+\pi^- \oplus p$ cross section
(expected from Regge to be $\sim$ln$s^{\sim -1.25}$) there is more rapidity available for proton dissociation at 1960 GeV, the beam
rapidities being 6.87 and 7.64 while the detector extends to $\eta$ = 5.9 in both cases. We observe that the ratio is lower
in the region of the $f_2(1270)$ than it is below 1 GeV, expected to be dominated by S-wave. We also find that the mean
$p_T(\pi^+\pi^-)$ has a minimum in the $f_2(1270)$ region, and rises abruptly at 1.5 GeV.

\begin{figure}[ht]
\centering

\subfigure[Differential cross section $d\sigma/dM$ for two particles, assumed to be $\pi^+\pi^-$, in the stated kinematic region, 
between two rapidity gaps $\Delta y >$ 4.6, at $\sqrt{s}$ = 900 GeV (red) 
and 1960 GeV (black).]{%
\includegraphics[width=70mm]{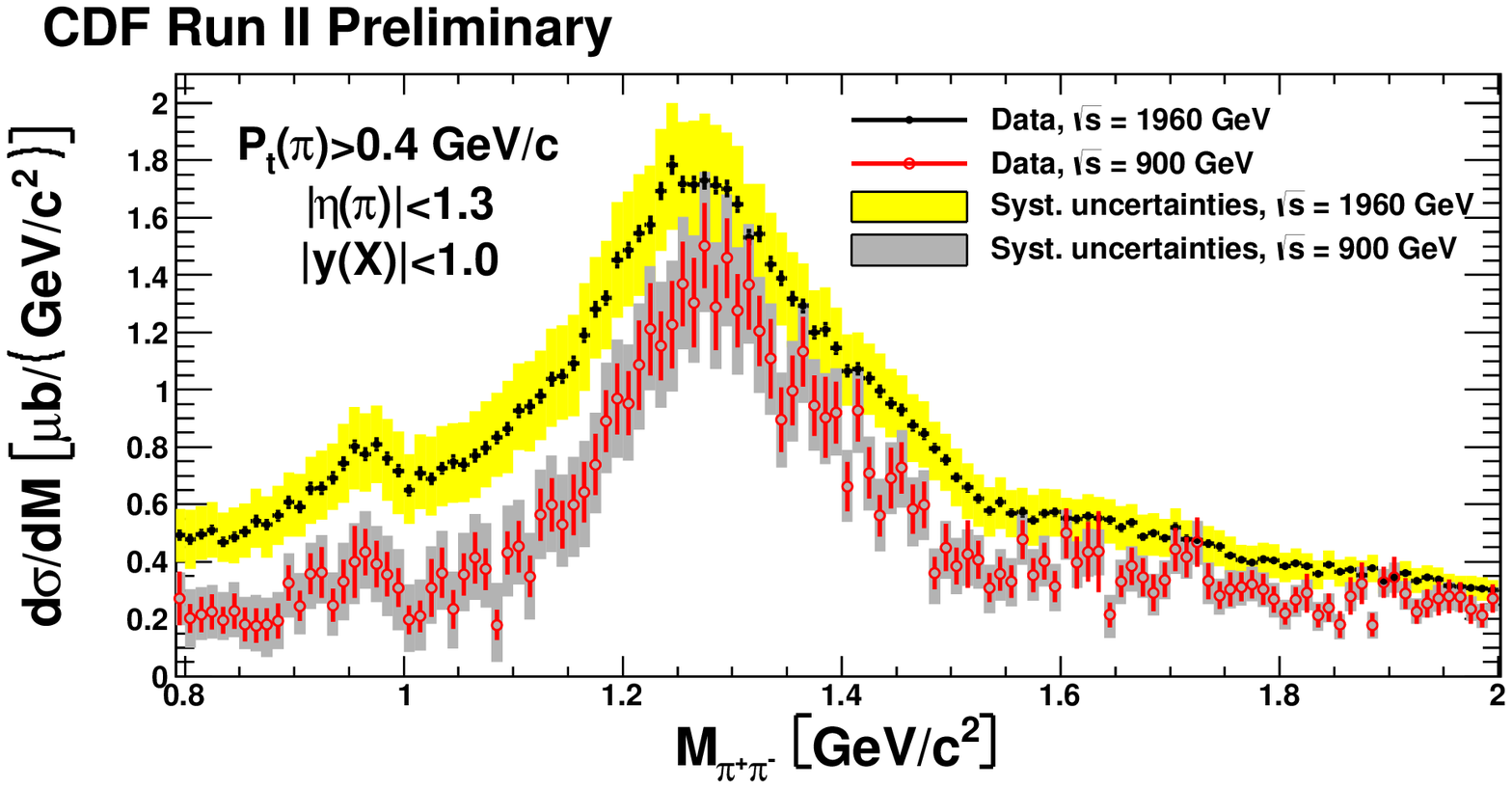}}
\label{fig:Mass0820_1960900}
\quad
\subfigure[Ratio of cross sections $d\sigma/dM$ at $\sqrt{s}$ = 1960 GeV 
and 900 GeV as a function of mass. In both cases rapidity gaps extend to $\eta_{max}$ = 5.9, and p-dissociation is
included.]{%
\includegraphics[width=70mm]{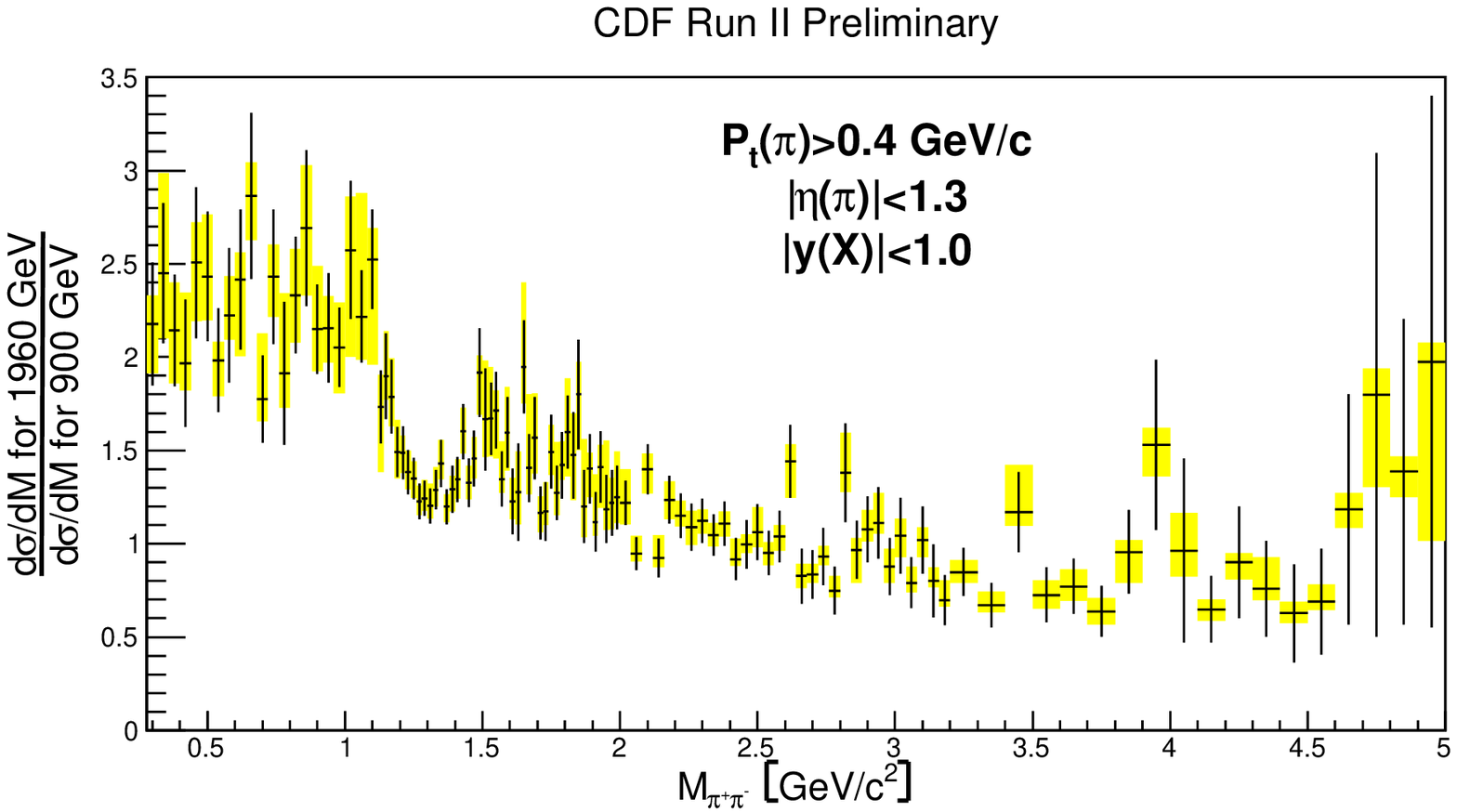}}
\label{fig:ratioplot}

\end{figure}

%\begin{figure}
%\centering
%\includegraphics[width=80mm]{Ratio.eps}
%\caption{}
%\label{fig:ratioplot}
%\end{figure}

We previously observed~\cite{cdfchic} exclusive $\chi_{c}^0$ production in the mode $J/\psi (\rightarrow \mu^+\mu^-) + \gamma$, but could
not distinguish the three $\chi_c$ states. The $\pi^+\pi^-$ and $K^+K^-$ channels have larger branching
fractions and enough resolution to separate the $\chi_{c}$ states. We do not see significant signals in this data,
and give upper limits (90\% C.L.) on $d\sigma/dy|_{y=0}(\chi_{c0})$ = 21.4$\pm$4.2(syst.)nb (in $\pi^+\pi^-$) and 18.9$\pm$3.8(syst.)nb
(in $K^+K^-$). This implies that $<25$\% of the $J/\psi+\gamma$ events were $\chi_{c0}(3415)$. Even though the $\chi_{c2}(3556)$
may have a much smaller production cross section its branching fraction is 17$\times$ larger.

We studied the cos $\theta^*$ distributions of the $\pi^+$ in the $X$-frame relative to the incoming $p$-direction.
The data are consistent with isotropy up to 1.5 GeV, above which they become progressively more forward-backward peaked.
Isotropy is expected if any polarization at production is washed out after integration over the unseen
protons or $p*$-dissociations. 

The ``Durham" and ``Cracow" groups ~\cite{durham,cracow} have predicted the differential cross section with the same cuts as 
Fig.~\ref{fig:Mass_Corr_1960}, but with no dissociation. Theoretical uncertainties in the region $\sim 3 < M < 4$ GeV are about
$^{\times 3}_{\div 3}$, but the data are within these uncertainties.

%We have data, and plan to analyse, also $X = K^+K^-, K^0_S K^0_S, \rho \rho, \eta\eta, \eta'\eta$, etc. There are many other
%studies possible when $M(X)$ is higher, e.g. looking for jets, double parton scattering, etc.

%\section{Acknowledgments}

We acknowledge funding from the U.S. Dept. of Energy and many other sources, see Ref.~\cite{cdfgg}.

\begin{footnotesize}
% IF YOU DO NOT USE BIBTEX, USE THE FOLLOWING SAMPLE SCHEME FOR THE REFERENCES
% ----------------------------------------------------------------------------

\end{footnotesize}
\end{document}